\documentclass[reprint,aps,pra,floatfix,longbibliography,onecolumn]{revtex4-2}
\usepackage{hyperref}
\usepackage{amsmath,amssymb}
\usepackage{graphicx,color}

\def\abs#1{{\lvert#1\rvert}}
\def\vec#1{{\boldsymbol#1}}
\let\Re\relax\DeclareMathOperator{\Re}{Re}
\let\Im\relax\DeclareMathOperator{\Im}{Im}
\DeclareMathOperator{\sgn}{sgn}
\DeclareMathOperator{\erf}{erf}

\begin{document}

\title{Particle and pair spectra for strongly correlated Fermi gases:
  A real-frequency solver}
\author{Tilman Enss}
\affiliation{Institut f\"ur Theoretische Physik, Universit\"at
  Heidelberg, D-69120 Heidelberg, Germany}

\begin{abstract}
  The strongly attractive Fermi gas in the BCS-BEC crossover is
  efficiently described in terms of coupled fermions and fermion
  pairs, or molecules.  We compute the spectral functions of both
  fermions and pairs in the normal state near the superfluid
  transition using a Keldysh formulation in real frequency.  The
  mutual influence between fermions and pairs is captured by solving
  the self-consistent Luttinger-Ward equations: these include both the
  damping of fermions by scattering off dressed pairs, as well as the
  decay of pair states by dissociation into two dressed fermions.  The
  pair spectra encode contact correlations between fermions and form
  the basis for computing dynamical response functions and transport
  properties.
\end{abstract}

\date{\today}
\maketitle

\section{Introduction}

Strongly correlated Fermi gases are ubiquitous in nature and appear in
very diverse physical realization ranging from ultracold atomic gases
\cite{bloch2008, giorgini2008} to dilute nuclear matter
\cite{strinati2018}.  In their theoretical description, however,
universality provides a framework to reveal the common features of
these systems. Recent experimental advances in atomic spectroscopy
have brought the dynamical properties and even transport within reach
of high-precision measurements \cite{li2024}.

In dilute, yet strongly attractive Fermi gases pair fluctuations play
a dominant role throughout the BCS-BEC crossover \cite{zwerger2012}.
They describe not only the condensation of fermion pairs in the
low-temperature superfluid state, but virtual pair fluctuations also
strongly renormalize the properties of the normal state above the
superfluid transition temperature $T_c$ \cite{nozieres1985,
  sademelo1993, perali2002}.  Pair fluctuations alone, however, are
not sufficient, and density (particle-hole) fluctuations lead to
a substantial reduction of $T_c$ at weak coupling \cite{gorkov1961}.
Also at resonant scattering in the unitary regime \cite{bloch2008} the
value of $T_c\simeq0.16T_F$ from experiment \cite{ku2012} and quantum
Monte Carlo \cite{burovski2006} is very well reproduced in field
theoretic approaches based on the \emph{self-consistent} T-matrix, or
Luttinger-Ward theory \cite{haussmann2007, hanai2014, frank2018,
  pini2019} (for related 
self-consistent GW approaches see \cite{yeh2022}).
This approach includes particle-hole fluctuations in the fermion and
pair self-energies, which are computed self-consistently at one-loop
order with fully dressed propagator lines.  Diagrammatic Monte Carlo
results confirm that further multi-loop contributions modify the
density equation of state by less than 10\% even at $T_c$
\cite{vanhoucke2012}.

In these approaches, the coupled self-consistent equations for
fermions and pairs have been solved numerically in imaginary
(Matsubara) frequency or time.  This is computationally convenient
because convergence properties are well understood.  However,
real-frequency spectra can only be obtained by analytical
continuation, which is mathematically ill-defined and requires
exponential precision in imaginary frequency to obtain reliable
real-frequency data.  Exponential precision, however, is not
achievable in numerical self-consistent solutions.  We propose,
therefore, to solve the self-consistent equations directly in real
frequency, which circumvents analytical continuation.  We present an
algorithm that computes fermion spectra and, for the first time,
self-consistent pair spectra reliably even with standard numerical
precision.  As we explain below, the main idea is to represent the
fermion and pair self-energies as slowly varying functions
interpolated on a real-frequency and momentum grid, then use
analytical integration between grid points to obtain highly accurate
spectra that capture also sharp spectral features much narrower than
the grid spacing.

This paper is structured as follows.  In section~\ref{sec:kel} we
introduce the Keldysh formulation of the strongly correlated Fermi gas
in equilibrium.  The self-consistent solution in real frequency is
developed in Sec.~\ref{sec:sc}.  In section~\ref{sec:spec} we present
the resulting fermion and pair spectra for the strongly correlated
three-dimensional Fermi gas in the BCS-BEC crossover.  We conclude in
Sec.~\ref{sec:con} and discuss how these results can form the basis
for future self-consistent computations of dynamical response
functions and transport directly in real frequency.

\section{Fermi gas model in Keldysh formulation}
\label{sec:kel}

\subsection{Attractive Fermi gas}

We consider a two-component Fermi gas in three dimensions, which is
described by the Hamiltonian
\begin{equation}
  \label{eq:Ham}
  H = \sum_{\sigma} \int d\vec r\, \psi_\sigma^\dagger(\vec r)
  \bigl[-\frac{\hbar^2\nabla^2}{2m} - \mu_\sigma\bigr] \psi_\sigma(\vec r)
  + g_0 \int d\vec r\, \psi_\uparrow^\dagger(\vec r)
  \psi_\downarrow^\dagger(\vec r) \psi_\downarrow(\vec r)
  \psi_\uparrow(\vec r).
\end{equation}
Here, $\psi_\sigma(\vec r)$ denotes the field operator for a fermion
of species $\sigma=\,\uparrow,\downarrow$ and mass $m$ at chemical
potential $\mu_\sigma$.  The second term represents an attractive contact
interaction between unlike fermions of bare strength $g_0<0$.  The
contact interaction needs to be regularized at short distance in two
and higher dimensions, and in three dimensions it is related to the
low-energy $s$-wave scattering length $a$ via
\begin{equation}
  \label{eq:scattlen}
  \frac1{g_0} = \frac m{4\pi\hbar^2a} - \frac{m\Lambda}{2\pi^2\hbar^2}
\end{equation}
in the presence of a large-wavenumber cutoff $\Lambda$.  The
attractive interaction tends to form pairs of fermions, and for
positive scattering length $a>0$ there exists a bound state of two
fermions at a binding energy of
\begin{equation}
  \label{eq:binding}
  E_b = \frac{\hbar^2}{ma^2} > 0.
\end{equation}
Even when the attraction is too weak to form a bound state at negative
scattering length $a<0$, virtual pair fluctuations play an important
role.  It is therefore natural to introduce a local pair field
\begin{align}
  \label{eq:pair}
  \Delta(\vec r)
  & = g_0 \psi_\downarrow(\vec r) \psi_\uparrow(\vec r),
  & \Delta_{\vec q}
  & = g_0 \int_{\vec k} \psi_{\vec q-\vec k\downarrow} \psi_{\vec k\uparrow}
\end{align}
in real or momentum space, respectively (the short-hand notation
$\int_{\vec k} \equiv \int d\vec k/(2\pi)^d$).  After a
Hubbard-Stratonovich transformation one obtains the following
Fermi-Bose action in terms of both fermion and pair degrees of freedom
\cite{sademelo1993, sachdev1999, enss2012crit},
\begin{align}
  \label{eq:1}
  S = \int d\vec r \int_0^\beta d\tau \left\{ \sum_\sigma
  \psi_\sigma^*\left(\partial_\tau -
  \frac{\nabla^2}{2m}-\mu_\sigma\right) \psi_\sigma
  -\frac1{g_0}\abs\Delta^2-\psi_\uparrow^* \psi_\downarrow^* \Delta -
  \Delta^* \psi_\downarrow \psi_\uparrow \right\},
\end{align}
where $\tau$ denotes imaginary time (we use units where $\hbar=1$ from
now on). Alternatively, one can start from a two-channel Hamiltonian
for fermions and molecules in the broad resonance limit \cite{bloch2008}.

\subsection{Keldysh technique in equilibrium}

The model of coupled fermions and pairs leads to a dressing of both,
which is reflected in their renormalized spectra.  We will use the
Keldysh formulation \cite{kamenev2011} in equilibrium in order to
compute these spectra in real frequency.  Although in principle the
equilibrium spectra could be obtained from Matsubara Green functions
\cite{abrikosov1975} by analytical continuation, this procedure is
mathematically ill-defined and error prone for noisy numerical data.
By using the Keldysh formulation we avoid the need for analytical
continuation and obtain reliable spectra directly in real frequency.

The bare retarded fermion Green function of spin component $\sigma$,
\begin{equation}
  \label{eq:greenf0}
  G_{\sigma 0}^R(\vec p,\varepsilon)
  = \frac1{\varepsilon+i0+\mu_\sigma-\varepsilon_p},
\end{equation}
encodes noninteracting particles with dispersion relation
$\varepsilon_p = p^2/2m$ and a bare spectral function
$A_{\sigma0}(\vec p,\varepsilon) = -(1/\pi)\Im G_{\sigma0}^R(\vec
p,\varepsilon) = \delta(\varepsilon+\mu_\sigma-\varepsilon_p)$.  In
the presence of interactions the bare Green function turns into the
fully dressed Green function with self-energy
$\Sigma_\sigma^R(\vec p,\varepsilon)$ via the Dyson equation,
\begin{equation}
  \label{eq:greenf}
  G_\sigma^R(\vec p,\varepsilon)
  = \frac1{\varepsilon+i0+\mu_\sigma-\varepsilon_p -
    \Sigma_\sigma^R(\vec p,\varepsilon)}
   = -i\int_0^\infty dt\,e^{i(\varepsilon+i0) t}\langle \{\psi_{\vec
     p\sigma}(t), \psi_{\vec p\sigma}^\dagger(0)\} \rangle.
\end{equation}
The bosonic Green functions that represents fermion pairs is
analogously given by (with subscript $p$ for pairs)
\begin{equation}
  \label{eq:greenb}
  G_p^R(\vec q,\omega)
  = \frac1{g_0^{-1} - \Sigma_p^R(\vec q,\omega)}
  = -i\int_0^\infty dt\,e^{i(\omega+i0) t}\langle [\Delta_{\vec q}(t),
  \Delta_{\vec q}^\dagger(0)] \rangle
\end{equation}
with bare coupling $g_0$ and bosonic self-energy $\Sigma_p^R(\vec
q,\omega)$.

\begin{figure}[t!]
  \centering
  \includegraphics[width=.8\linewidth]{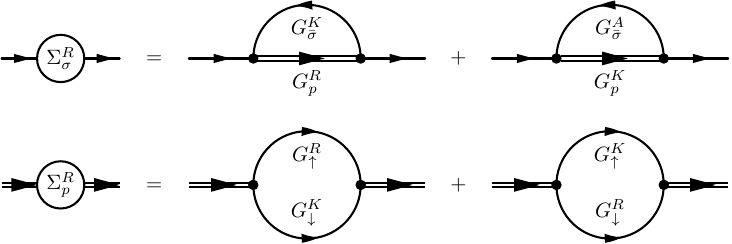}
  \caption{Feynman diagrams for the fermionic and pair self-energies
    in Keldysh formulation, with $G^R$ (retarded), $G^A$ (advanced) and $G^K$
    (Keldysh) propagators.  All propagator lines are bold and
    represent fully dressed fermions (single) and pairs (double
    lines).}
  \label{fig:schematic}
\end{figure}

In our model fermions can scatter off real or virtual pairs and
acquire a fermionic self-energy that is given by \cite{kamenev2011,
  kawamura2020} (here $\bar\sigma = -\sigma$ denotes the other fermion
species)
\begin{equation}
  \label{eq:sigmafkel}
  \Sigma_\sigma^R(\vec p,\varepsilon)
  = -\frac i2 \int_{\vec p',\varepsilon'}
  \bigl[ G_p^R(\vec p+\vec p',\varepsilon+\varepsilon')
  G_{\bar\sigma}^K(\vec p',\varepsilon') +
  G_p^K(\vec p+\vec p',\varepsilon+\varepsilon')
  G_{\bar\sigma}^A(\vec p',\varepsilon') \bigr],
\end{equation}
see Fig.~\ref{fig:schematic}.  While the retarded and advanced Green
functions $G^R(\vec p,\varepsilon) = [G^A(\vec p,\varepsilon)]^*$
represent only the spectrum, the Keldysh components
$G^K(\vec p,\varepsilon)$ represent both the spectrum and the
occupation number with the statistical factor in equilibrium:
\begin{align}
  \label{eq:keldysh}
  G_\sigma^K(\vec p,\varepsilon)
  & = -i \tanh(\beta\varepsilon/2) A_\sigma(\vec p,\varepsilon), \\
  G_p^K(\vec q,\omega)
  & = -i \coth(\beta\omega/2) A_p(\vec q,\omega),
\end{align}
with full spectral functions
$A_\sigma(\vec p,\varepsilon) = -(1/\pi)\Im G_\sigma^R(\vec
p,\varepsilon)$ and
$A_p(\vec q,\omega) = -(1/\pi)\Im G_p^R(\vec q,\omega)$.  The
occupation factors can be rewritten in terms of the Fermi and Bose
functions as $\tanh(\beta x/2) = 1-2f(x)$ with
$f(x)=\tfrac12(1-\tanh(\beta x/2))$ and as
$\coth(\beta x/2) = 1+2b(x)$ with $b(x)=\tfrac12(\coth(\beta x/2)-1)$.
In this way, the retarded fermionic self-energy is expressed as
\begin{equation}
  \label{eq:sigmaf}
  \Sigma_\sigma^R(\vec p,\varepsilon)
  = \int_{\vec p',\varepsilon'}
  \bigl[ G_p^R(\vec p+\vec p',\varepsilon+\varepsilon')
  f(\varepsilon') A_{\bar\sigma}(\vec p',\varepsilon') -
  b(\varepsilon+\varepsilon') A_p(\vec p+\vec p',\varepsilon+\varepsilon')
  G_{\bar\sigma}^A(\vec p',\varepsilon') \bigr],
\end{equation}
where a contribution independent of occupation vanishes by
analyticity.  For the imaginary part of the self-energy the occupation
factors are combined with a product of spectral functions as
\begin{equation}
  \label{eq:imsigmaf}
  \Im \Sigma_\sigma^R(\vec p,\varepsilon)
  = -\pi \int_{\vec p',\varepsilon'}
  [ f(\varepsilon') + b(\varepsilon+\varepsilon') ]
  A_p(\vec p+\vec p',\varepsilon+\varepsilon')
  A_{\bar\sigma}(\vec p',\varepsilon').
\end{equation}
Once the imaginary part has been computed, the real part can be
obtained by the Kramers-Kronig relation
\begin{equation}
  \label{eq:KK}
  \Re \Sigma^R(\vec p,\varepsilon)
  = \int \frac{d\varepsilon'}{\pi} \,
  \mathcal P \frac{\Im\Sigma^R(\vec p,\varepsilon')}
  {\varepsilon'-\varepsilon},
\end{equation}
which involves an integral over the principal value $\mathcal P$.

The bosonic self-energy, in turn, arises from dissociation of a pair into
individual fermions and is computed as the particle-particle bubble
diagram (see Fig.~\ref{fig:schematic}),
\begin{align}
  \label{eq:sigmabkel}
  \Sigma_p^R(\vec q,\omega)
  & = \frac i2 \int_{\vec p,\varepsilon}
  \bigl[ G_\uparrow^R(\vec q-\vec p,\omega-\varepsilon)
  G_\downarrow^K(\vec p,\varepsilon) +
  G_\uparrow^K(\vec q-\vec p,\omega-\varepsilon)
  G_\downarrow^R(\vec p,\varepsilon) \bigr] \\
  & = \int_{\vec p,\varepsilon}
  \bigl[ G_\uparrow^R(\vec q-\vec p,\omega-\varepsilon)
  [\tfrac12 - f(\varepsilon)] A_\downarrow(\vec p,\varepsilon)
  + [\tfrac12 - f(\omega-\varepsilon)]
  A_\uparrow(\vec q-\vec p,\omega-\varepsilon)
  G_\downarrow^R(\vec p,\varepsilon) \bigr]. \notag
\end{align}
Both terms can be combined after a change of variables to yield
\begin{equation}
  \label{eq:sigmab}
  \Im\Sigma_p^R(\vec q,\omega)
  = -\pi\int_{\vec p,\varepsilon} [1-2f(\varepsilon)]
  A_\uparrow(\vec p,\varepsilon) A_\downarrow(\vec q-\vec p,\omega-\varepsilon).
\end{equation}
Causality implies that the imaginary part of the fermionic Green
function is always negative,
$\Im G_\sigma^R(\vec p,\varepsilon) < 0 \; \forall \varepsilon$, while the
imaginary part of the bosonic Green function changes sign at
$\omega=0$, $\Im G_p^R(\vec q,\omega) \sgn(\omega) < 0$.  The same
holds for the sign of the imaginary parts of the fermionic and bosonic
self-energies, which follows from their definitions \eqref{eq:sigmaf}
and \eqref{eq:sigmab}.  Equations \eqref{eq:greenf}, \eqref{eq:greenb},
\eqref{eq:sigmaf} and \eqref{eq:sigmab} form a
closed set of coupled integral equations for the fermion and pair
Green functions.  This particular set of equations corresponds to the
self-consistent Luttinger-Ward approach \cite{baym1962, haussmann1993,
  haussmann2007}.  In the following we present a new method for their
numerical solution in real frequency.

The Keldysh technique introduced so far applies to general polarized
Fermi gas with $\mu_\uparrow\neq\mu_\downarrow$. In this work we will
start by presenting the solution for the case of a balanced
(unpolarized) gas with $\mu_\uparrow=\mu_\downarrow=\mu$ and
$G_\uparrow = G_\downarrow \equiv G_\sigma$.

\subsection{Quantum virial expansion}

Since we are interested in the strongly correlated Fermi gas at large
scattering length $\abs{a}$, the interaction strength is not a good
expansion parameter.  Instead, in the high-temperature normal state
one can perform a quantum virial expansion in the fermionic fugacity
\begin{align}
  \label{eq:z}
  z = \exp(\beta\mu)
\end{align}
as the small parameter, where $\beta=1/(k_BT)$ denotes the inverse
temperature and we work henceforth in units where $k_B\equiv1$.  In
the high-temperature virial expansion we can already identify spectral
features that will be important reference points in the discussion of
the low-temperature spectra below.  When pairing is important the
\emph{pair fugacity}
\begin{align}
  \label{eq:zp}
  z_p = \exp(\beta\mu_p) = \exp(\beta[2\mu+E_b]) = z^2e^{\beta E_b}
\end{align}
controls the strength of pair contributions with pair chemical
potential $\mu_p = 2\mu+E_b$.  One can distinguish the
fermion-dominated regime $z_p < z$ from the pair-dominated regime
$z_p > z$ \cite{fujii2023bulk}.  In the expressions for the self-energy
\eqref{eq:sigmaf} and \eqref{eq:sigmab}, the Fermi function is
expanded for small fugacity as
$f(x-\mu) = ze^{-\beta x} + \mathcal O(z^2)$ and the Bose function as
$b(x-\mu_p) = z_p e^{-\beta x} + \mathcal O(z_p^2) = z^2e^{\beta E_b}
e^{-\beta x} + \mathcal O(z^4)$.

To zeroth order in fugacity, i.e., in vacuum, the fermion self-energy
vanishes and the bosonic self-energy is known analytically as
\begin{align}
  \label{eq:sigmab0}
  \Sigma_{p0}^R(\vec q,\omega)
  & = \int_{\vec p}^\Lambda G_{\sigma0}^R(\vec q-\vec p,\omega+\mu-\varepsilon_p) \\
  & = \frac{m}{4\pi}\sqrt{-m(\omega+2\mu-\tfrac12\varepsilon_q+i0)} -
    \frac{m\Lambda}{2\pi^2} \notag
\end{align}
for large cutoff $\Lambda\to\infty$.  The cutoff term in the
definition of the bare coupling $g_0$ \eqref{eq:scattlen} cancels that
in the self-energy to yield the cutoff independent pair Green function
\cite{nozieres1985}
\begin{equation}
  \label{eq:greenb0}
  G_{p0}^R(\vec q,\omega)
  = \frac{4\pi/m}{a^{-1} -
    \sqrt{-m(\omega+2\mu-\tfrac12\varepsilon_q+i0)}}.
\end{equation}
The corresponding pair spectral function reads
\begin{equation}
  \label{eq:specb0}
  A_{p0}(\vec q,\omega)
  = \frac{4\pi}{m^{3/2}} \Bigl[ 2\sqrt{E_b}\delta(s_p+E_b)\Theta(a) +
  \frac1\pi \frac{\sqrt{s_p} \Theta(s_p)}{s_p+1/(ma^2)}
  \Bigr]_{s_p=\omega+2\mu-\varepsilon_q/2}
\end{equation}
in terms of the pair spectral parameter $s_p=\omega+2\mu-\varepsilon_q/2$,
which measures the energy from the onset of the scattering continuum
at $s_p=0$.  The pair spectrum exhibits a scattering continuum for $s_p>0$
from the square root branch cut, and for positive $a>0$ there is
additionally the pair bound state at $s_p=-E_b$ with pair dispersion
$\omega_q = q^2/(2M)$ at twice the fermion mass, $M=2m$.  Note that
the pair spectrum in vacuum is still Galilean invariant, i.e., it
depends only on the combination $s_p$ and not on
$\omega$ or $\vec q$ separately.  This will no longer be the case at
finite density, as we shall see below.

At finite density the Fermi distribution has to be included in the
bosonic self-energy even when using bare fermion propagators $G_{\sigma0}$,
and the bosonic self-energy for bare fermions reads
\begin{align}
  \label{eq:sigmab1}
  \Sigma_p^{(1)R}(\vec q,\omega)
  & = \int_{\vec p} [1-2f(\varepsilon_p-\mu)]
    G_{\sigma0}^R(\vec q-\vec p,\omega+\mu-\varepsilon_p), \\
  \label{eq:sigmab1im}
  \Im \Sigma_p^{(1)R}(\vec q,\omega)
  & = \frac{m}{4\pi} \Bigl[ -\sqrt{ms_p} +
    \frac{2mT}{q} \ln
    \frac{1-f(\omega/2+\sqrt{\varepsilon_qs_p/2})}
    {1-f(\omega/2-\sqrt{\varepsilon_qs_p/2})} \Bigr] \Theta(s_p).
\end{align}
While the real part is not known analytically, it is easily obtained
by numerical Kramers-Kronig transformation \eqref{eq:KK} since the
finite-temperature correction (second term) decays exponentially for
large frequency $\omega\gg T$.  In the limit $q\to0$ the imaginary
part simplifies to
$\Im\Sigma_p^{(1)R}(\vec q=0,\omega) = -(m/4\pi) \sqrt{ms_p}
\tanh(\beta\omega/4) \Theta(s_p)$.  Both in this expression and in
Eq.~\eqref{eq:sigmab1im} the sign of the imaginary part changes at
$\omega=0$ as required by causality.  As is clear from
Eq.~\eqref{eq:sigmab1im}, the self-energy at nonzero density depends
no longer only on $s_p$ but also on $\omega$ or $q$ separately and the
pair propagator in medium does not have a Galilean invariant form.

As the density is further increased the attractive Fermi gas undergoes
a phase transition into a superfluid state.  This occurs when pairs
can be excited at zero momentum $q=0$ and zero energy $\omega=0$, as
given by the Thouless criterion $[G_p^R(\vec q=0,\omega=0)]^{-1} = 0$
\cite{nozieres1985}.  With bare fermions this is equivalent to
$a^{-1}=(2/\pi)\int_0^\infty ds\, \frac{\sqrt{ms}}{s-2\mu}
f(s/2-\mu)$.  At weak coupling $a\to0^-$ it yields the critical
temperature $(\beta\mu)_c^{-1} = (8e^{\gamma-2}/\pi)\exp(\pi/2k_Fa)$,
but at stronger coupling no analytical expression is known.  At
unitarity, the non-self-consistent calculation with bare propagators
yields $(\beta\mu)_c\approx 1.5$, while experiment \cite{ku2012} and
self-consistent Luttinger-Ward theory \cite{haussmann2007} yield a
value of $(\beta\mu)_c\approx 2.5$, corresponding to
$T_c/T_F\approx0.16$.

\section{Self-consistent solution in real frequency}
\label{sec:sc}

At low temperatures in the normal state, $T>T_c$, one can reach
$z\gtrsim1$ and the virial expansion does not converge.  Instead, we
will now present a real-frequency solver for the coupled integral
equations
(\ref{eq:greenf},\ref{eq:greenb},\ref{eq:sigmaf},\ref{eq:sigmab}).  In
continuous time and space, one can choose a grid of frequencies
$\varepsilon_i$ to sample the full domain
$-\infty < \varepsilon < \infty$ and a grid of radial momenta $p_j$
for $0\leq p<\infty$.  The challenge of a straightforward numerical
solution is that the spectral function $A(\vec p,\varepsilon)$ can
have very narrow $\delta$ peaks at large frequencies or momenta even
in a strongly correlated system, making it difficult to resolve these
peaks on a grid. Such narrow peaks, however, often arise as simple
poles where the denominator has a zero crossing and is well described
by a linear approximation of the \emph{denominator} between grid
points. Hence, we propose an ``inverse'' integrator that interpolates
the denominator of the Green function.  Such an approximation becomes
accurate if the self-energy $\Sigma(\vec p,\varepsilon)$ varies only
slowly with $\vec p$ and $\varepsilon$. Using a linear interpolation
of the self-energy between grid points, the value of the frequency or
momentum integral of the Green function can be computed analytically,
depending on the value of $\Sigma$ on the adjacent grid points.
Explicitly, we obtain with $x=(p^2-p_j^2)/(p_{j+1}^2-p_j^2)$
\begin{align}
  \label{eq:logint}
  \int_{p_j}^{p_{j+1}} dp\,p\, G(p,\varepsilon)
  = \frac{p_{j+1}^2-p_j^2}2 \int_0^1 \frac{dx}{a+bx}
%  = \frac{p_{j+1}^2-p_j^2}{2b} \ln \frac{a+b}a
  = \frac{p_{j+1}^2-p_j^2}2 \frac{\ln (G_{j+1}^{-1}G_j)}{G_{j+1}^{-1}-G_j^{-1}}\,,
\end{align}
where $a=G^{-1}(p_j,\varepsilon)$ and
$a+b=G^{-1}(p_{j+1},\varepsilon)$ are given in terms of the
self-energy at the grid points.  An analogous formula applies for
integration over frequency $\varepsilon$ between grid points.  This
integral is exact for noninteracting particles and it is a good
approximation for slowly varying $\Sigma$; note that there is no need
that $\Sigma$ be small, hence the validity extends to strong coupling
beyond the virial expansion. The numerical results below show that
$\Sigma$ changes in frequency roughly on the scale of the temperature
$T$ and in momentum over $\sqrt{2mT}$, so in practice it is sufficient
to use a frequency grid $\varepsilon=-100T\dotsc100T$ with equidistant
spacing $\Delta\varepsilon=0.5T$ and a momentum grid
$p=0\dotsc10\sqrt{2mT}$ with spacing $\Delta p=0.2\sqrt{2mT}$.  For
the lowest temperatures in our study, $T/T_F=0.16$, this corresponds
to a self-energy grid spacing $\Delta\varepsilon=0.08\varepsilon_F$
and $\Delta p=0.08k_F$.

The explicit formula for the one-loop renormalization of the fermion
and pair spectra, which has the form of a convolution integral
\eqref{eq:imsigmaf}, can be written as
\begin{align}
  \label{eq:imsfmom}
  \Im \Sigma_\sigma^R(\vec p,\varepsilon)
  & = -\pi \int_{-\infty}^\infty d\varepsilon'\,
  [ f(\varepsilon') + b(\varepsilon+\varepsilon') ]
  \frac1{4\pi^2p} \int_0^\infty dq\, q A_p(q,\varepsilon+\varepsilon')
  \int_{|p-q|}^{p+q} dp'\, p' A_{\bar\sigma}(p',\varepsilon') \notag
  \\
  & \equiv \text{Convolution}_\sigma[\Sigma_p,\Sigma_{\bar\sigma}].
\end{align}
Here, the momentum integral has been expressed as an integral over the
bosonic radial momentum $q$ and the fermionic radial momentum $p'$,
weighted by $\Theta(|p-q|<p'<p+q)/(4\pi^2p)$. The principal function
of the $p'$ integral is computed first using formula
\eqref{eq:logint}, which takes $\mathcal O(N)$ time for $N$
momentum-frequency grid points.  While the spectral function can have
$\delta$ peaks, its principal function has at most steps of unit
height, such that the error from a discretization $\Delta p$ of the
momentum integral is at most $\mathcal O(\Delta p)$.  The subsequent
$q,\varepsilon'$ integration for every $p,\varepsilon$ takes
$\mathcal O(N^2)$ time for the complete self-energy, which is the most
time-intensive computational step. Finally, the Kramers-Kronig
transformation \eqref{eq:KK} is employed to obtain the real part of
the self-energy; this ensures exact analyticity of the self-energy
even if $\Im\Sigma$ is computed approximately.

At high temperature or weak coupling the fermionic self-energy follows
approximately the result of perturbation theory, which has a
square-root nonanalyticity, but the absolute magnitude of the
self-energy is small.  Conversely, at low temperature and strong
coupling the self-energy is much larger (of order $T$) but at the same
time the functional form is smoothed by an imaginary part of order
$T$.  As a result, the slope of the fermionic self-energy with respect
to energy, $|d\Sigma_\sigma/d\varepsilon|$, is bounded as
$\mathcal O(1)$ for all energies and momenta.  This is confirmed
numerically for the self-energies shown in Fig.~\ref{fig:specufg} below.
Hence, the maximum discretization error in $\Sigma_\sigma$ is bounded
as $\mathcal O(\Delta\varepsilon)$ and by refining the grid spacing
the discretization error can be systematically reduced.

The frequency integral is weighted by the fermionic and bosonic
occupation factors.  While the fermionic occupation is always positive
and smooth, weighting by the bosonic occupation requires care as
$b(\omega) \simeq \frac T\omega-\frac12$ for $|\omega|\ll T$.  In the
Keldysh formulation, however, the Bose function appears always in
combination with a bosonic spectral function $A_p(\omega)$, which
changes sign at $\omega=0$ and scales linearly in $\omega$ for
$|\omega|\ll T$.  The product of bosonic spectrum and occupation,
therefore, is continuous and well-behaved near $\omega=0$. One can
easily check that also the combination
$f(\varepsilon') + b(\varepsilon+\varepsilon')$ in
Eq.~\eqref{eq:imsfmom} changes sign when $\varepsilon+\varepsilon'=0$
such that the
$\Im\Sigma_\sigma^R(p,\varepsilon)<0\;\forall\varepsilon$, as required
by causality.

As we will see below, the fermionic self-energy $\Im\Sigma_\sigma$
scales asymptotically as $\varepsilon^{-1/2}$ for large frequencies.
This tail is important in the Kramers-Kronig transformation in order
to accurately obtain the real part $\Re\Sigma_\sigma$ near the Fermi
surface, which determines the shift of the spectral lines.  We include
the asymptotic tail beyond the maximum grid frequency analytically in
the Kramers-Kronig transformation.  In the convolution integral,
instead, the high-frequency parts are suppressed by the Fermi and Bose
distributions and give only a small contribution.

Analogously, the bosonic self-energy is computed
explicitly as
\begin{align}
  \label{eq:imsbmom}
  \Im \Sigma_p^R(\vec q,\omega)
  & = -\pi \int_{-\infty}^\infty d\varepsilon\,
  [ 1-2f(\varepsilon) ]
  \frac1{4\pi^2q} \int_0^\infty dp\, p A_\uparrow(p,\omega-\varepsilon)
  \int_{|p-q|}^{p+q} dp'\, p' A_\downarrow(p',\varepsilon) \notag \\
  & \equiv \text{Convolution}_p[\Sigma_\uparrow,\Sigma_\downarrow].
\end{align}
The set of self-consistent integral equations can be solved by
iteration.  As initial condition we choose bare fermions with
$\Sigma_\sigma^{(0)}(p,\varepsilon)\equiv-i0$ (in practice we use
$-10^{-8}iT$).  Each iteration consists of two steps:
\begin{enumerate}
\item Compute the new pair self-energy from Eq.~\eqref{eq:imsbmom} and
  numerical Kramers-Kronig transformation \eqref{eq:KK} for the real part,
  \begin{align}
    \label{eq:iterbos}
    \Sigma_p^{(i+1)} = \text{Convolution}_p[\Sigma_\uparrow^{(i)},\Sigma_\downarrow^{(i)}].
  \end{align}
  In the first iteration for $\Sigma_p^{(1)}$ we use the analytical
  formula \eqref{eq:sigmab0} for accuracy.
\item Compute the new fermion self-energy from Eqs.~\eqref{eq:imsfmom}
  and \eqref{eq:KK},
  \begin{align}
    \label{eq:iterferm}
    \Sigma_\sigma^{(i+1)}
    = \alpha\, \text{Convolution}_\sigma[\Sigma_p^{(i+1)},\Sigma_{\bar\sigma}^{(i)}] +
    (1-\alpha) \Sigma_\sigma^{(i)}.    
  \end{align}
  A convergence enhancement factor $0<\alpha\leq1$ can accelerate
  convergence, and the converged result for $\alpha<1$ is the same
  self-consistent solution as for $\alpha=1$.
\end{enumerate}
Convergence is typically reached after a dozen steps (a bit slower
near $T_c$); we denote the converged self-consistent solution as
$\Sigma^{(\infty)}$ and $G^{(\infty)}$.  In practice, convergence is
fastest if one uses the fully converged solution for a given parameter
value $\mu$ as initial condition for a new calculation at a
neighboring parameter value $\mu+\delta\mu$.

\begin{figure}[t!]
  \centering
  \includegraphics[width=\linewidth]{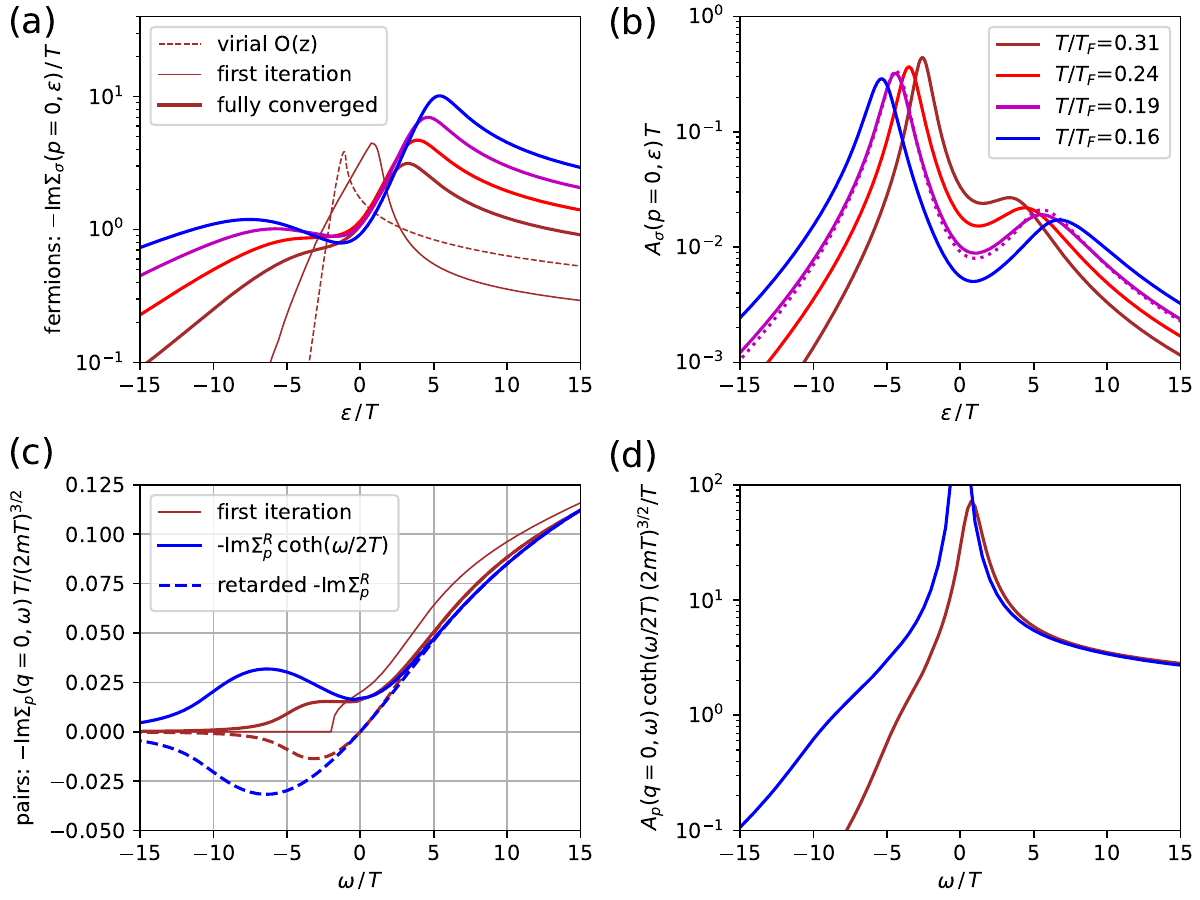}
  \caption{Line spectra of fermions and pairs at unitarity: real
    frequency dependence at zero momentum. Temperatures are in the
    normal state approaching $T_c$: $T/T_F=0.31$ ($\beta\mu=1.0$,
    brown), $T/T_F=0.24$ ($\beta\mu=1.5$, red), $T/T_F=0.19$
    ($\beta\mu=2.0$, magenta), and $T/T_F=0.16$ ($\beta\mu=2.5$,
    blue). (a) The fermion self-energy shows a single peak in the
    virial expansion (dashed) and a renormalized single peak in the
    first LW iteration $\Sigma_\sigma^{(1)}$ (thin line); instead, the
    fully converged self-energy $\Sigma_\sigma^{(\infty)}$ (thick
    line) develops a two-peak structure which grows more prominent as
    the temperature is lowered. (b) The fermion spectrum shows a
    two-peak structure as a precursor to the Bogoliubov spectrum.  The
    comparison with spectral data of Ref.~\cite{johansen2024} (dotted)
    shows good agreement.  (c) The pair self-energy has a zero
    crossing at $\omega=0$ by causality (dashed); the spectral weight
    at negative frequencies is enhanced at lower temperature. The
    Keldysh component $-\Im\Sigma_p(q=0,\omega)\coth(\omega/2T)$
    (solid) remains positive at all frequencies and regular around
    $\omega=0$. (d) The pair spectrum has a single, asymmetric peak
    near threshold that becomes broader and more pronounced at lower
    temperature.}
  \label{fig:sigmaufg}
\end{figure}

\section{Luttinger-Ward spectra}
\label{sec:spec}

The procedure in Sec.~\ref{sec:sc} produces self-consistent
solutions for single-particle and pair Green functions, $G_\sigma^R$ and
$G_p^R$, in real frequency.  A crucial feature of this Luttinger-Ward
approach is that the results are thermodynamically consistent
\cite{baym1962, enss2012crit}: the density, for instance, is obtained
identically either from the grand potential by thermodynamic
derivative, or by loop integration over the single-particle Green
function as
\begin{align}
  \label{eq:2}
  n_\sigma
  & = \int_{-\infty}^\infty d\varepsilon\, f(\varepsilon) g_\sigma(\varepsilon),
  & g_\sigma(\varepsilon)
  & = \int \frac{d^3p}{(2\pi)^3}\, A_\sigma(p,\varepsilon), \\
  \mathcal C
  & = m^2 \int_{-\infty}^\infty d\omega\, b(\omega) g_p(\omega),
  & g_p(\omega)
  & = \int \frac{d^3q}{(2\pi)^3}\, A_p(q,\omega),
\end{align}
where we have defined the single-component fermion density $n_\sigma$
and density of states $g_\sigma(\varepsilon)$.  In the second line,
the local pair density, or contact $\mathcal C$, is defined via an
integral over the pair density of states $g_p(\omega)$, with an
additional mass factor $m^2$ by convention.  The thermodynamic results
for the fermion density and the contact (pair) density agree with
previous self-consistent Luttinger-Ward computations in imaginary
frequency \cite{haussmann2007}. In order to showcase our method, we
choose applications to two distinct physical regimes of the strongly
correlated Fermi gas, (i) the unitary regime at the scattering
resonance ($a^{-1}=0$) where spectra are dominated by many-body
BCS-type pairing ($z_p\leq z^2$), and (ii) the BEC regime
($(k_Fa)^{-1}\simeq1$) where two-body binding has a substantial
effect on many-body properties ($z_p\gg z^2$).

\begin{figure}[t!]
  \centering
  \includegraphics[width=\linewidth]{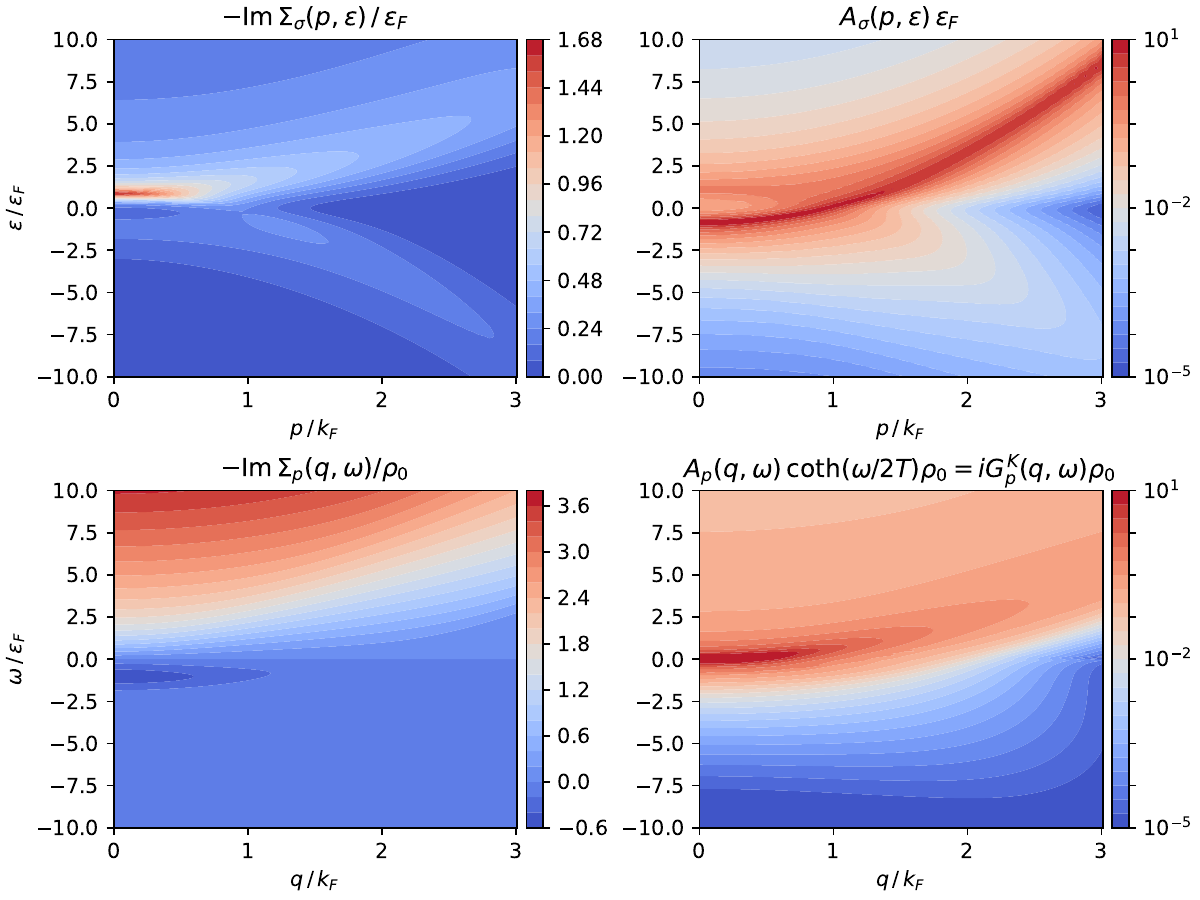}
  \caption{Luttinger-Ward self-energies and spectral functions at unitarity
    $1/a=0$ and temperature $T/T_F=0.16$ ($\beta\mu=2.5$).  (a) Fermion
    self-energy.  (b) The fermion spectral function shows a band splitting
    around the Fermi level $\varepsilon=0$ and a slight suppression of
    spectral weight also near $k_F$.  (c) Pair
    self-energy.  (d) The pair spectral function weighted by the Bose
    factor (Keldysh component) is positive and strongly peaked at the
    threshold of the scattering continuum.  Pair functions are given in
    units of the zero-temperature density of states
    $\rho_0=g_\sigma^{(0)}=mk_F/2\pi^2$.}
  \label{fig:specufg}
\end{figure}

\subsection{Particle and pair spectra at unitarity}

Figures \ref{fig:sigmaufg} and \ref{fig:specufg} show the resulting
self-energies and spectra for both fermions and pairs in the unitary
regime ($\beta E_b=0$).  First figure~\ref{fig:sigmaufg} presents the
line spectra for the frequency dependence at vanishing momentum, which
already exhibits many qualitative features of the full spectrum in
Fig.~\ref{fig:specufg}.  For $\beta\mu=1$ in the quantum degenerate
regime, corresponding to $T/T_F\approx0.3$, the high-temperature
virial expansion \eqref{eq:sigmaf1virialU} to first order in fugacity
predicts a single narrow peak in the imaginary part of the self-energy
(brown dashed line in Fig.~\ref{fig:sigmaufg}(a)).  The first
iteration of the Luttinger-Ward scheme, computed with bare fermions,
includes contributions to arbitrary order in fugacity and displays
already a much broader single peak for this large value of fugacity,
$z=e^1$.  The full self-consistent solution, instead, shows a
qualitatively different behavior with a two-peak structure, which
becomes more pronounced as the chemical potential is increased to
$\beta\mu=1.5,\,2.0,\,2.5$, corresponding to temperatures approaching
$T/T_F\simeq0.16$ just above $T_c$ (blue curve).  Correspondingly, the
fermion spectrum in Fig.~\ref{fig:sigmaufg}(b) develops a two-peak
structure as a precursor to the Bogoliubov spectrum as the temperature
is lowered.

The imaginary part of the pair self-energy in
Fig.~\ref{fig:sigmaufg}(c) arises from dissociation of pairs into
individual fermions: it has a square-root branch cut representing the
scattering continuum for large frequencies.  As a bosonic function it
must change sign at $\omega=0$ (dashed curves).  When weighted with
the Bose factor, $-\Im\Sigma_p(\omega)\coth(\omega/2T)>0$ is regular
at $\omega=0$ and positive for all frequencies (solid curves).  Note
that the fully converged self-consistent solution (thick lines)
contains substantially more spectral weight at smaller frequencies
$\omega<0$ than the first iteration (thin line), in particular at
lower temperature (blue).  Finally, the pair spectral function in
Fig.~\ref{fig:sigmaufg}(d) weighted with the Bose factor (the Keldysh
component \eqref{eq:keldysh}) is positive and exhibits a single large
peak for the onset of the scattering continuum, which decays for large
frequencies as $\omega^{-1/2}$.  This slow decay, in turn, determines
also the decay of the fermionic self-energy as $\varepsilon^{-1/2}$
for large frequencies.

The full spectra at $T=0.16T_F$ slightly above $T_c$ are shown in
Fig.~\ref{fig:specufg}.  The fermion self-energy has a broad upward
branch starting at $\varepsilon\sim \varepsilon_{\vec p}/2$, which arises from
combining with another low-momentum fermion into a pair state
(molecule-hole continuum), and a steeper downward branch
$\varepsilon\sim -\varepsilon_{\vec p}$, which arises from combining with
another fermion into a low-momentum pair.  The imaginary part of the
self-energy determines the decay rate (inverse lifetime) of the
corresponding fermion states, which near $T_c$ is substantial
(comparable to $\varepsilon_F$).  These spectral features are
reflected in the fermion spectral function Fig.~\ref{fig:specufg}(b),
which shows a band splitting around the Fermi level $\varepsilon=0$
and the appearance of Bogoliubov shadow bands already above $T_c$.
Note that the fermionic self-energy and spectral function clearly do
not follow a single-parameter scaling in terms of the spectral
parameter $s=\varepsilon+\mu-\varepsilon_{\vec p}$ alone.

The pair self-energy in Fig.~\ref{fig:specufg}(c) clearly exhibits the
scattering continuum for $\omega \gtrsim \varepsilon_{\vec q}/2-2\mu$.
In the self-consistent solution, where a pair can dissociate into
dressed fermion states, the substantial broadening of the fermions
shifts the threshold of the scattering continuum to lower frequency
with respect to the non-selfconsistent solution.  Finally, the full
pair spectrum in Fig.~\ref{fig:specufg}(d) shows the clear threshold
of the scattering continuum as well as an additional downward branch
that arises from the dressed fermions.

\begin{figure}[t!]
  \centering
  \includegraphics[width=\linewidth]{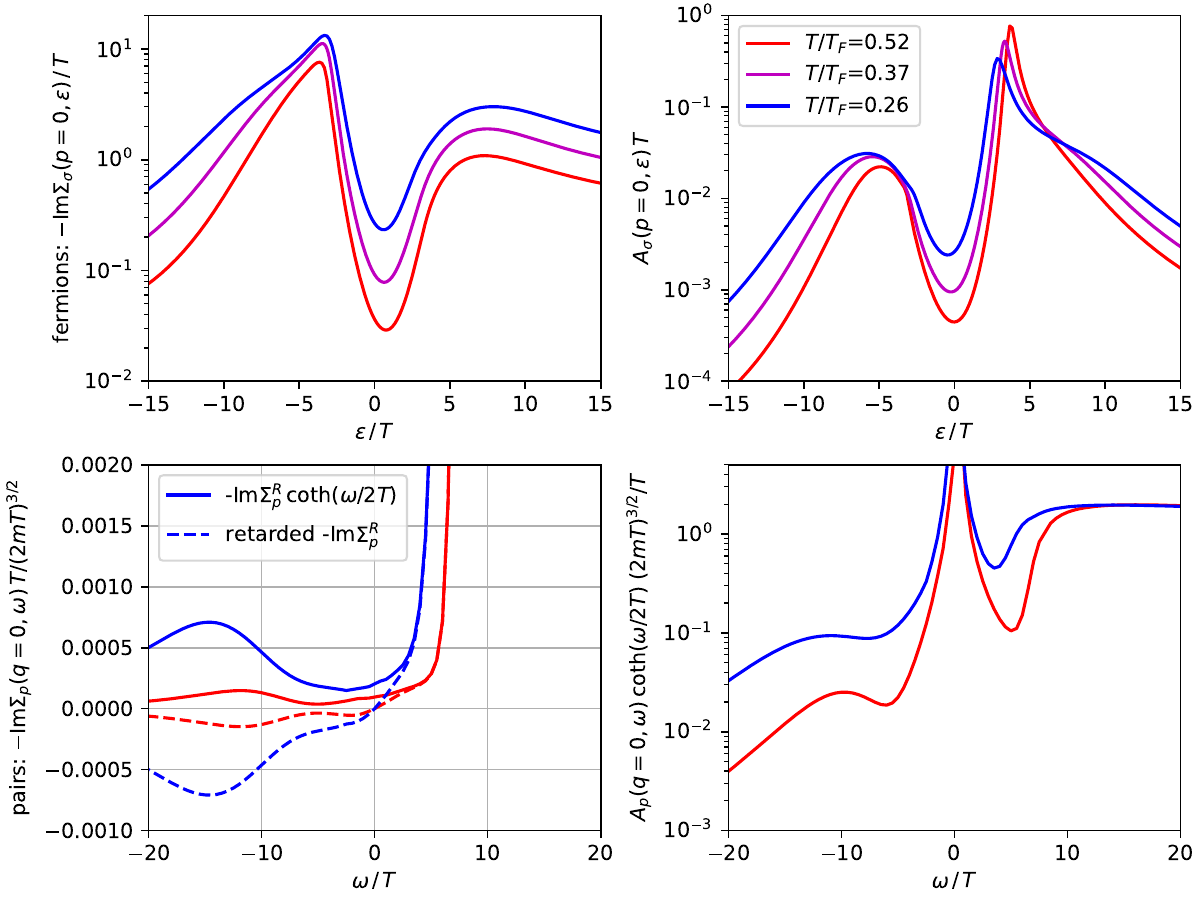}
  \caption{Line spectra of fermions and pairs in the BEC regime
    ($\beta E_b=8$): real frequency dependence at zero momentum.
    Temperatures are in the normal state approaching $T_c$:
    $T/T_F=0.52$ ($\beta\mu=-4$, red), $T/T_F=0.37$ ($\beta\mu=-3.75$,
    magenta), and $T/T_F=0.26$ ($\beta\mu=-3.5$, blue).  (a) The
    fermion self-energy shows a much more pronounced two-peak
    structure than at unitarity.  (b) The fermion spectrum similarly
    shows a two-peak structure as a precursor to the Bogoliubov
    spectrum.  (c) The
    pair self-energy has a zero crossing at $\omega=0$ by causality
    (dashed); the spectral weight at negative frequencies is enhanced
    at lower temperature. The Keldysh component
    $-\Im\Sigma_p(q=0,\omega)\coth(\omega/2T)$ (solid) remains
    positive at all frequencies and regular around $\omega=0$.  (d)
    The pair spectrum has a triple peak structure: a large bound-state
    peak near $\omega=0$, the scattering continuum for positive
    $\omega$ separated by a gap from the bound state, and a small peak
    at negative frequency.}
  \label{fig:sigmabec}
\end{figure}

\begin{figure}[t!]
  \centering
  \includegraphics[width=\linewidth]{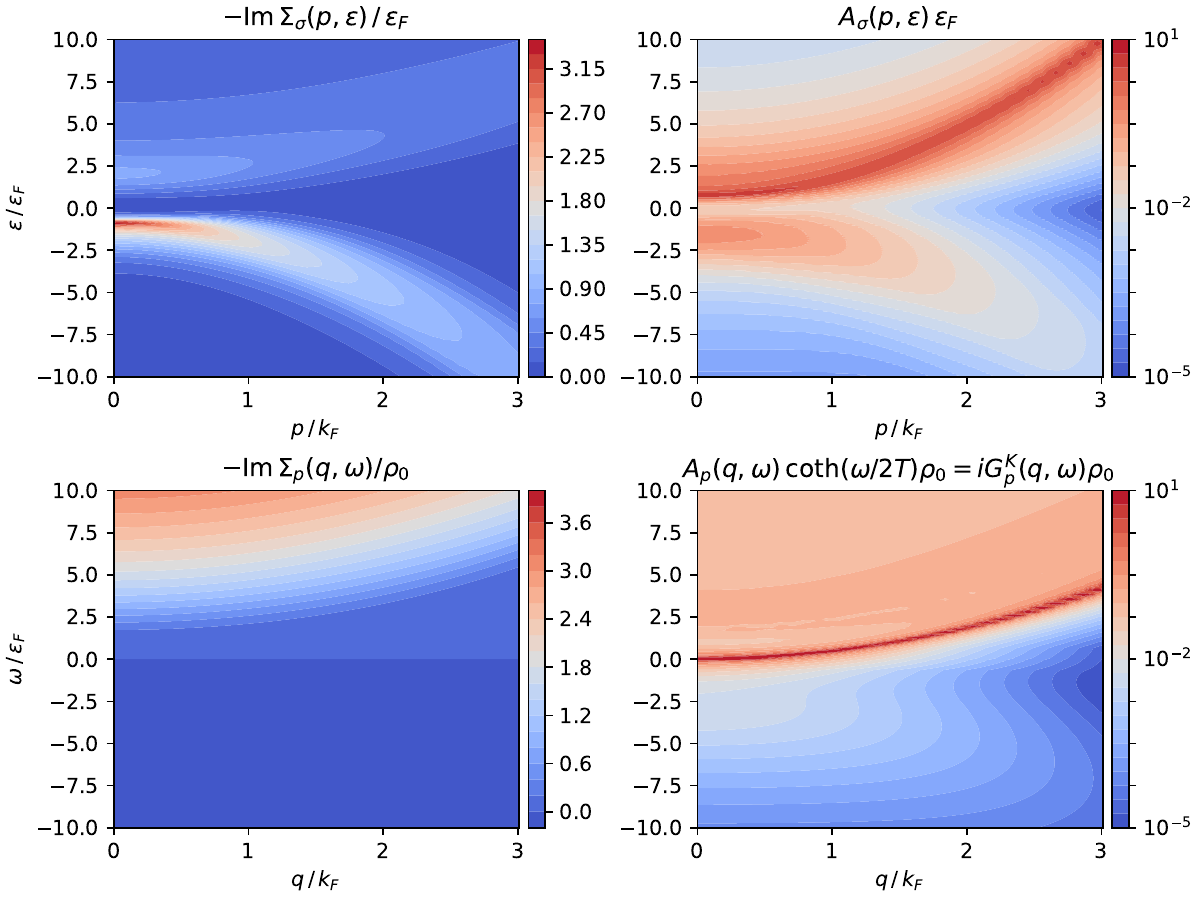}
  \caption{Luttinger-Ward self-energies and spectra in the BEC regime
    $1/(k_Fa)=1$ at $T/T_F=0.26$ ($\beta\mu=-3.5$) slightly above
    $T_c$. (a) The fermion self-energy $\Sigma_\sigma(p,\varepsilon)$
    shows two branches, and fermions scatter most strongly on the
    lower branch. (b) The fermion spectral function
    $A_\sigma(p,\varepsilon)$ shows a clear gap between the two
    branches around the Fermi level $\varepsilon=0$ and has most
    spectral weight concentrated on the upper branch. (c) The pair
    self-energy $\Sigma_p(q,\omega)$ shows the two-particle scattering
    continuum. (d) The weighted pair spectral function
    $A_p(q,\omega)\coth(\omega/2T)=iG_p^K$ (the pair Keldysh function)
    shows a strong bound-state branch separated by the binding energy
    $E_b=2E_F$ from the pair continuum, as well as a weak branch
    bending down.}
  \label{fig:specbec}
\end{figure}

\subsection{Particle and pair spectra in BEC regime}

In the BEC regime the fermion line spectra in Figs.~\ref{fig:sigmabec}
and \ref{fig:specbec} show many of the same qualitative features, such
as upward and downward branches, as in the unitary regime; however,
the splitting between the two branches in the fermionic spectrum is
now much larger, $\approx2\abs\mu>0$, and grows with momentum, as in the
strong-binding limit of the BCS dispersion relation \cite{murthy2018}.
The pair self-energy is dominated by the scattering continuum but has
again significant weight at negative frequency that arises from the
downward branch of the dressed fermions.  Finally, the pair spectral
function (Keldysh component) in Figs.~\ref{fig:sigmabec}(d) and
\ref{fig:specbec}(d) exhibits a \emph{three-peak} structure: the large
bound-state peak near $\omega=0$ becomes broader for lower
temperature; the scattering continuum is separated from the bound
state by a gap comparable to the binding energy $E_b$; and in addition
there is a downward branch at negative frequencies.

\begin{figure}[t!]
  \centering
  \includegraphics[width=.5\linewidth]{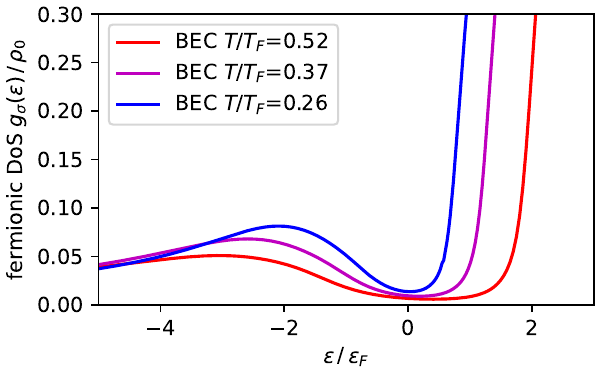}
  \caption{Fermionic density of states (DoS) $g_\sigma(\varepsilon)$
    vs frequency $\varepsilon$ on the BEC side $\beta E_b=8$
    ($1/k_Fa\simeq1$).  The DoS is strongly suppressed in a region of
    width $2\abs\mu$ around the chemical potential. The DoS is given
    in units of the ideal Fermi gas DoS at zero temperature,
    $\rho_0=g_\sigma^{(0)}=mk_F/(2\pi^2)$.}
  \label{fig:dos}
\end{figure}

The fermion dispersion exhibits qualitative differences between the
unitary regime, where it resembles the BCS-type dispersion relation
with minimum gap at nonzero wave vector $k_*\approx k_F$, and the BEC
regime, where the gap is present at all $k$ and reaches a minimum at
$k=0$.  This qualitative change between the two regimes is also
apparent in density of states (DoS).  While at unitarity the density
of states is only slightly suppressed near the Fermi level
$\varepsilon=0$ above $T_c$ (cf.\ Fig.~\ref{fig:specufg}(b)), in the
BEC regime the gap is clearly developed already in the normal state,
cf.~Fig.~\ref{fig:dos}, but instead it becomes narrower (in units of
$\varepsilon_F$) toward lower temperature.

\section{Discussion}
\label{sec:con}

The real-frequency solver presented in this work circumvents the
long-standing problem of analytical continuation by computing a
self-consistent solution directly in the Keldysh spectral
representation.  This gives access to the dynamical properties of
single particles, which agree with previous results where available
\cite{haussmann2009}; at strong coupling they show a substantial
renormalization of spectra compared to the virial expansion, and in
particular the self-consistent algorithm allows us to access the
low-temperature regime $\beta\mu>1.5$ at unitarity, which is
unattainable in bare perturbation theory.  The self-consistent
solution in the Luttinger-Ward framework ensures thermodynamic
consistency and the exact fulfillment of Tan relations, as well as
scale invariance in the unitary case even for approximate solutions
\cite{enss2012crit}.  In particular, existing thermodynamic results
(e.g., $\mu(n)$) obtained in imaginary frequency can be used as input
for the real-frequency computation.  On the technical level, in the
Keldysh formulation the divergence of the bosonic occupation at zero
frequency is compensated by the smallness of the bosonic spectral
function to yield a well-defined frequency integral, but it can still
have sharp peaks from long-lived excitations.  Our algorithm treats
these efficiently by interpolation of the self-energy, which is a
slowly varying function between grid points.  The accuracy of the
spectra is confirmed by comparing with results for finer grids
($\Delta\varepsilon=0.25T$) and with the spectral data of
Ref.~\cite{johansen2024}.  By construction, the resulting Green
functions satisfy the requirements of analyticity and causality.  The
Luttinger-Ward spectra with their subtle three-peak structure
(Fig.~\ref{fig:sigmabec}) can serve as a benchmark and as a prior for
the numerical reconstruction of imaginary-time quantum Monte Carlo
data.

The real-frequency solver can immediately be applied to imbalanced
(polarized) Fermi gases with $\mu_\uparrow\neq\mu_\downarrow$, in fact
the equations in Sec.~\ref{sec:kel} are already written for this
general case.  This will allow one to extend self-consistent
ground-state polaron spectra \cite{schmidt2011} to finite
temperature.  Furthermore, the self-consistent thermodynamics in the
symmetry-broken, superfluid state has been found in the balanced
\cite{haussmann2007} and imbalanced Fermi gas \cite{frank2018}, and it
will be worthwhile to extend the real-frequency solver to this case in
order to obtain the corresponding excitation spectra.  Another
important extension will be to the two-dimensional Fermi gas
\cite{levinsen2015strongly}, which always admits a pair bound state with
$E_b>0$ and is therefore covered by our algorithm for the BEC regime;
this will allow for a self-consistent computation of pairing spectra
\cite{bauer2014, murthy2018} and the dynamical quantum scale anomaly
\cite{hofmann2012, murthy2019}.

Another very interesting extension, which is subject of ongoing
work, is to compute dynamical correlation and response functions,
which define, e.g., transport coefficients such as shear viscosity
\cite{massignan2005, enss2011}, bulk viscosity \cite{dusling2013,
  enss2019bulk, nishida2019, hofmann2020, fujii2023bulk}, thermal
conductivity \cite{braby2010, frank2020quantum}, and spin diffusivity
\cite{enss2012spin, enss2019spin}.  As the frequency dependent
transport coefficients depend on the slope in frequency of a bosonic
spectral function, the real-frequency solver should yield improved
self-consistent predictions both at zero and finite frequency.
Finally, the Keldysh formulation can describe genuine out-of-equilibrium
dynamics where the fluctuation-dissipation relation \eqref{eq:keldysh}
is no longer satisfied \cite{kamenev2011}, and it will be interesting
to find self-consistent solutions for the transient evolution after a
quantum quench.

After completion of this work two other studies appeared which compute
spectral functions in real frequency using Fourier transforms
\cite{johansen2024} and spectral representations \cite{dizer2023}.

We acknowledge discussions with E.~Dizer, B.~Frank, K.~Fujii, E.~Gull,
J.~Lang, G.~M\"oller, J.~Pawlowski, F.~Werner, D.~Zgid, and
W.~Zwerger, and thank J.~Lang for sharing spectral data.  This work is
supported by the Deutsche Forschungsgemeinschaft (German Research
Foundation), project-ID 273811115 (SFB1225 ISOQUANT) and under
Germany’s Excellence Strategy EXC2181/1-390900948 (the Heidelberg
STRUCTURES Excellence Cluster).

\appendix

\section{Leading order self-energies in the virial expansion}

Beyond the vacuum limit, the fermionic self-energy arises at order
$\mathcal O(z)$.  To leading order, one can replace the fermionic
occupation $f(x-\mu)=ze^{-\beta x}+\mathcal O(z^2)$ and neglect the
bosonic occupation $b(y-2\mu)=\mathcal O(z^2)$.  For the first-order
self-energy it is thus sufficient to use the zeroth-order Green
functions $G_{p0}^R$ and
$A_{\sigma0}(\vec p',\varepsilon') =
\delta(\varepsilon'+\mu-\varepsilon_{p'})$, such that
Eq.~\eqref{eq:sigmaf} simplifies to
\begin{align}
  \label{eq:sigmaf1}
  \Sigma_\sigma^{(1)R}(\vec p,\varepsilon=\varepsilon_{\vec p}-\mu+s,a^{-1})
  & = z \int_{\vec p'} e^{-\beta\varepsilon_{p'}}\,
    G_{p0}^R(\vec p+\vec p',\varepsilon+\varepsilon_{p'}-\mu) \\
  & = z \frac{4\pi}m \int_{\vec p'} \frac{e^{-\beta\varepsilon_{p'}}}
    {a^{-1}-\sqrt{-m(s+\tfrac12 \varepsilon_{\vec p-\vec p'} +
    i0)}}. \notag
\end{align}
The fermion spectral parameter $s=\varepsilon+\mu-\varepsilon_{\vec p}$
denotes the shift away from the fermion shell, which is located at
$s=0$.  Analytical expressions for the first-order self-energy
$\Sigma_\sigma^{(1)R}(\vec p,s,a^{-1})$ are available if only one of
the three parameters $p$, $s$, and $a^{-1}$ is nonzero:
\begin{align}
  \Sigma_\sigma^{(1)R}(\vec p,0,a^{-1}\to0)
  \label{eq:sigmaf1virial}
  & = \frac{2zT}{\sqrt\pi} \Bigl[ -\frac{4a^{-1}}{\sqrt{2mT}} \,
    \frac{F_D(p/\sqrt{2mT})}{p/\sqrt{2mT}} -i
    \frac{\erf(p/\sqrt{2mT})}{p/\sqrt{2mT}} \Bigr], \\
  \Sigma_\sigma^{(1)R}(0,s,0)
  \label{eq:sigmaf1virialU}
  & = -i\frac{2zT}{\sqrt\pi} U(\tfrac12,0,2s/T+i0), \\
  \Sigma_\sigma^{(1)R}(0,0,a^{-1})
  & = i\frac{8zT}{\pi^2}\,\frac{a^{-2}}{2mT}\, G\bigl[ ((1,3/2,2),()),
    ((3/2,2),()), i\frac{\sqrt{2mT}}{2a^{-1}}, 1/2 \bigr], \\
  \Sigma_\sigma^{(1)R}(0,-E_b,a^{-1})
  & = -i\frac{2zT}{\sqrt\pi} G\bigl[ ((),(3/2)),((0,1),()),2/a^2mT
    \bigr] + \text{(real part)}.
\end{align}
Here, $F_D$ denotes the Dawson, $U$ the hypergeometric, and $G$ the
Meijer function.  The unitary on-shell self-energy
\eqref{eq:sigmaf1virial} was given in \cite{dusling2013}, and
specifically $\Sigma_\sigma^{(1)R}(0,0,0)=-i4zT/\pi$; the subsequent
results are to our knowledge new.  Additionally, the bound-state
contribution for all argument values is given analytically to order
$\mathcal O(z)$ by
\begin{align}
  \Im \Sigma_\sigma^{(1)bR}(\vec p,s,a^{-1}>0)
  & = -2zT \Theta(a) \Theta(-s-E_b) \sqrt{2E_b/T} \,
    \frac{e^{-(\bar p-\bar p')^2}-e^{-(\bar p+\bar p')^2}}{4\bar p}
\end{align}
with $\bar p=p/\sqrt{2mT}$ and $\bar p'=\sqrt{-2(s+E_b)/T}$, and the
last ratio becomes $\bar p'\exp(-\bar p'^2)$ in the limit $\bar
p\to0$.  These analytical forms serve as a benchmark for the accuracy
of the numerical solution at small $z$.

\bibliography{all}

\end{document}